\documentstyle[pra,aps]{revtex}
\draft

\begin{document}

\font\Bbb= msbm10
\def\S{\mbox{\bf S}}
\def\eff{\mathrm{\scriptstyle eff}}
\def\mag{\mathrm{\scriptstyle mag}}
\def\N{{\it N}}
\def\T{{\it T}}
\def\I{{\it I}}

\twocolumn[\hsize\textwidth\columnwidth\hsize\csname@twocolumnfalse%
\endcsname

\title {Topology and Nematic Ordering II: Observable Critical Behavior}

\author{John~Toner} \address{ \centerline{IBM  T. J. Watson Research Center,
Yorktown Heights, NY  10598} }
\author{Paul~E.~Lammert}
%\address{Lab. de Physico-Chimie Th\'eorique,
\address{Ecole Sup\'erieure de Physique et de Chimie Industrielles,
10 rue Vauquelin, 75231 Paris Cedex 05, France}
\author{Daniel~S.~Rokhsar}
\address{Department of Physics, University of California, Berkeley, CA 94720}

\date{January 20, 1995}
\maketitle
\begin{abstract}
This paper is the second in a pair treating a new lattice model for
nematic media.  In addition to the familiar isotropic (\I) and
nematically ordered (\N) phases, the phase diagram established in the
previous paper (Paper I) contains a new, topologically ordered  phase (\T)
occuring at large suppression of topological defects and weak nematic
interactions.
This paper (Paper II) is concerned with the experimental signatures of
the proposed phase diagram.
Specific heat, light scattering and magnetic susceptibility near both
the \N/\T\ and \I/\T\ transitions are studied, and critical behavior
determined.
The singular dependences of the Frank constants $K_1$, $K_2$, $K_3$
and the dielectric tensor anisotropy $\Delta \epsilon$ on
temperature as $T \to T_{NT}^-$ are also found.
\end{abstract}
]

\section{Introduction and Results}
\label{results}

In the preceding paper\cite{part1} (hereafter referred to as ``Paper I''),
we showed that the nematic-to-isotropic transition need not proceed through
a single first order phase transition, contrary to long-held
belief\cite{de Gennes}.  In fact, we have shown that this disordering can
proceed by a pair of continuous transitions, with a novel intermediate
phase possessing non-trivial topological order.
The phase diagram discussed in Paper I (figure 3 there) is
expected to be generic for nematic materials since the gauge theory
from which it is derived depends only on the symmetry of nematic media.
The present paper is devoted to the calculation of
characteristic signatures of the phase transitions into and out of the
topologically ordered but physically isotropic phase \T\ discovered in
Paper I.
The quantities we study are the specific heat, light scattering, and
magnetic susceptibility near each transition.  We also calculate
the singular temperature dependences of the Frank constants
$K_1$, $K_2$, and $K_3$ as well as the dielectric tensor anisotropy
$\Delta \epsilon$.

An unusual feature of our model is the presence of two distinct, physically
isotropic phases, \T\ and \I, which are separated by a continuous
phase transition.  Unlike the familiar transition between liquid and gas
(which are also both isotropic fluids), there is no latent heat and no
critical endpoint between \T\ and \I; they cannot be smoothly interpolated.
The qualitative distinction between these two phases, both of which are
nematically disordered, is subtle: in the topologically ordered phase
there are few configurations with long nematic defects, and the locally
coarse grained director field can be unambiguously converted into a
non-singular vector field throughout the fluid.  The clearest demonstration
of this order would be the measurement of the energy per unit length
of a defect that is externally imposed by boundary conditions.
Such a direct demonstration of topological order, however, is likely to
be quite difficult.

Although one cannot use local probes to distinguish \T\ from \I\ deep
within these phases, the transition {\em between} them has very
characteristic and measurable properties.  Our prediction is that
under certain conditions, one should be able to observe critical
behavior by all conventional means (specific heat singularities,
light scattering, {\it etc.}) between two isotropic phases.
Observation of such critical behavior would support our scenario.
Verification of the scaling laws derived below would confirm the
universality classes identified in Paper I.

A less dramatic but also surprising prediction of Paper I is the
existence of a continuous phase transition between the nematically
ordered and the topologically ordered states.  According to conventional
wisdom based on Landau theory, the development of nematic order from any
isotropic state should occur as a first order phase transition.
As we have shown in Paper I, this need not be the case if the physically
isotropic phase has topological order, since then the transition can be
in the universality class of the three-dimensional Heisenberg
model.  Detailed predictions based on our model are presented below.

For convenient reference, we summarize our results here.
Detailed derivations are found in later sections.
Unless otherwise indicated, reduced temperature relative
to a particular transition is denoted by
$t \equiv (T - T_{\rm c})/T_{\rm c}$,
and the corresponding correlation length is denoted by $\xi$.
Subscripts $+$ and $-$ denote quantities pertaining above ($t > 0$)
and below ($t < 0$) $T_c$.
``Above'' always refers to the more disordered phase, so that the
isotropic phase is above the topologically ordered phase, which in
turn is above the nematically ordered state.

\vskip4pt
{\bf 1. Specific heat.}
The singular specific heat near the continuous transitions
into and out of the topologically ordered phase varies as
\begin{equation}
C(T) = A_\pm |t|^{-\alpha}.
\end{equation}
Near the \I/\T\ transition, the exponent $\alpha$ takes the (positive)
three-dimensional Ising model value\cite{exponents}
$\alpha_I = 0.1085 \pm 0.0075$, which implies
a divergent specific heat.  The amplitude ratio $A_+/A_-$ is inverted
relative to the usual Ising spin model, a result of the duality
transformation discussed in part I.
Near the \T/\N\ transition, $\alpha$ takes the three-dimensional Heisenberg
value\cite{exponents} $\alpha_{\rm H} = -0.130 \pm 0.021$,
resulting in a cusp.  The amplitude ratio $A_+/A_-$ takes the corresponding
universal value.

\vskip4pt
{\bf 2.  Polarized light scattering between isotropic states}.
Near the \I/\T\ transition, the light scattering intensity
$I_{xx}({\bf q},t)$ for incoming and scattered light with
parallel polarizations obeys the scaling law
\begin{equation}
I_{xx}({\bf q},t) = q^{-\alpha_I/\nu_I} f^{ITP}_{\pm}(q\xi)  \propto
\left\{ \begin{array}{ll}
t^{-\alpha_I}, &  \hskip0.1cm q\xi \ll 1; \\
q^{-\alpha_I/\nu_I}, & \hskip0.1cm q\xi \gg 1.
\end{array}
\right.
\label{light1}
\end{equation}
Here $\alpha_I$ and $\nu_I$ are the
three-dimensional Ising specific heat and correlation
length exponents\cite{exponents}
$\alpha_I = 0.1085 \pm 0.0075$,
$\nu_I = (2-\alpha_I)/3 = 0.6305 \pm 0.0025$,
respectively.  (The Ising correlation length $\xi$ diverges
as $t^{-\nu_I}$.)

\vskip4pt
{\bf 3. Depolarized light scattering between isotropic states}.
Near the \I/\T\ transition, the light scattering intensity
$I_{xy}({\bf q},t)$ for incoming and scattered beams with perpendicular
polarizations shows only a weak non-analyticity at
${\bf q} = 0$.
That is, $I_{xy}({\bf q}, t)$ can be decomposed into an analytic
and a singular part
\begin{equation}
I_{xy}^{{\rm sing}}({\bf q}, t) = q^{(1-\alpha_I) / \nu_I}
f^{ITD}_\pm(q\xi)
\propto
\left\{ \begin{array}{ll}
t^{1-\alpha_I}, &  \hskip0.1cm q\xi \ll 1; \\
q^{(1-\alpha_I)/ \nu_I}, & \hskip0.1cm q\xi \gg 1;
\end{array}
\right.
\label{light2}
\end{equation}
where $f^{ITD}_\pm$ is a universal function.
Note that the analytic contributions to the scattering dominate the
total scattering as $|{\bf q}| \to 0$, $t \to 0$.

The exponent in (\ref{light2}) is $(1-\alpha_I)/\nu_I = 1.414 \pm 0.006,$
which is greater than unity.  The singularity is therefore
merely a divergence in the {\em second} derivative $\partial^2
I_{xy}/\partial q^2$ of the scattering intensity.
%Such a weak singularity is difficult to detect and is therefore not
%a promising method of looking for the \I/\T\ transition.
This singularity is not a promising signature of
the \I/\T\ transition, since it is so weak.

\vskip4pt
{\bf 4. Depolarized light scattering near the nematic phase}.
Near the \N/\T\ transition, the depolarized light scattering intensities
$I_{xz}({\bf q},t)$ and $I_{yz}({\bf q},t)$ obey the scaling laws
\begin{equation}
I_{iz}({\bf q},t) = q^{2\eta_H-1} f^{NTD}_\pm (q\xi), \qquad i=x,y.
\label{light4}
\end{equation}
(The incoming polarization $\hat z$ is along the mean director.)
The very small anomalous dimension of the spin for the
three-dimensional Heisenberg model is $\eta_H = 0.02 \pm 0.01$.
Matching the form (\ref{light4}) onto the known small-wavenumber behavior of
$I_{iz}$ deep in the nematic and isotropic phases (for this purpose, \T
and \I\ behave identically) implies
%\addtocounter{equation}{1}
%\setbox0=\hbox{\hfill(\theequation\hbox{a})}
\begin{equation}
I_{iz}({\bf q},t) \propto
\left\{ \begin{array}{llr}
q^{-2} t^{\nu_H(1+2\eta_H)}, & \: t < 0,\: q\xi \ll 1; \\
q^{2\eta_H-1}, & \: q\xi \gg 1;  \\
t^{-\nu_H(1-2\eta_H)}, & \: t > 0,\: q\xi \ll 1.
\nonumber
\end{array}
\right.
\label{light3}
\end{equation}
Approaching the transition from the nematic side ($t < 0$), this
result implies a relationship between the critical behavior of the
Frank constants and the dielectric anisotropy $\Delta \epsilon$, since
\begin{equation}
I_{xz} = {c (\Delta \epsilon)^2 \over K_1 q_x^2 + K_3 q_z^2}
\label{For-1}
\end{equation}
and
\begin{equation}
I_{yz} = {c (\Delta \epsilon)^2 \over K_2 q_x^2 + K_3 q_z^2}\;,
\label{For-2}
\end{equation}
with some constant $c$ (we have taken $q_y = 0$).
The form  (\ref{light3}) then implies both that (i) the Frank constants
are asymptotically equal upon approaching $T_{NT}$ (due to the isotropy
of the scaling law) and that (ii) there is a particular relation between the
exponents governing the vanishing of the Frank constants and of $\Delta
\epsilon$ at the \N/\T\  transition.

These are, indeed, our next two conclusions.

\vskip4pt
{\bf 5.  Frank constants}.
The Frank constants approach a common value and vanish as
\begin{equation}
K_i \sim |t|^{\nu_H}, \qquad {\rm as}\quad T \to T_{NT}^-
\label{Frank-scale}
\end{equation}
at the \N/\T\  transition.  This can be tested by direct measurements of the
Frank constants, either by light scattering (see Item (5))  or by
Freedericksz instability measurements (see Item 8).

The approach of the ``splay'' Frank constant $K_1$ to equality with the
``bend'' and ``twist'' constants $K_2$ and $K_3$, however, is very slow.
Specifically, we expect
\begin{equation}
{K_1 - K_{2,3} \over K_{2,3}} \propto {1 \over |\ln(|t|)|}.
\label{twistshout}
\end{equation}
Experimentally, this behavior may look very much like $K_1$ being
proportional to $K_{2,3}$ with a constant prefactor of order unity.

\vskip4pt
{\bf 6. Dielectric anisotropy near the nematic phase.}
The dielectric anisotropy tensor $\Delta \epsilon$ vanishes like
\begin{equation}
\Delta \epsilon \sim |t|^{2\beta_H}, \qquad {\rm as}\quad T \to T_{NT}^-,
\end{equation}
where\cite{exponents} $\beta_H = 0.368 \pm 0.004$ is the Heisenberg
order parameter exponent.
This result for $\Delta \epsilon$ can be tested either through light
scattering as mentioned above, or by simple optical birefringence measurements.
Combining with our previous results (Items (4) and (5)) for the
Frank constants and $I_{iz}$ gives
\begin{equation}
I_{iz}({\bf q},t < 0) \propto q^{-2} |t|^{4 \beta_H -\nu_H}.
\end{equation}
With the exact relation $2 \beta = \nu (d - 2 + \eta)$ that is derived
by combining the Fisher and Rushbrooke scaling relations\cite{Ma},
we find eq.  (\ref{light3}) for $q\xi \ll 1$ (and $d=3$).

\vskip4pt
{\bf 7.  Magnetic response near the nematic phase.}
Near the \N/\T\ transition, the magnetic response of the system
is given by
\begin{equation}
M_ \alpha = \chi_I H_\alpha + \Delta \chi (t,H) \biggl( \hat{n}_\alpha
\hat{n}_\beta -{1 \over 3} \delta_{\alpha \beta} \biggr) H_\beta,
\label{magnetic-resp}
\end{equation}
where $\alpha$ denotes Cartesian components, {\bf n} is the
nematic director (always along $\bf H$ in the topologically ordered phase),
and the isotropic part of the susceptibility $\chi_I$
is analytic through the (continuous) \N/\T\  transition.
The {\em anisotropic} part of the susceptibility $\Delta \chi$,
on the other hand, obeys
\begin{equation}
\Delta \chi(t,H) = A |t|^{2\beta_H} f_\pm^\chi \left(
{H \over H_c |t|^{\phi_\chi}} \right),
\label{chi}
\end{equation}
where $A$ and $H_c$ are non-universal ({\it i.e.}, system-dependent)
constants.
The exponent is $\phi_\chi = \nu y_\tau/2 = 0.62\pm 0.01$, where
$y_\tau$ is the renormalization group eigenvalue of the ``spin-tensor
interaction\cite{Ma}'' for the ($n=3$) Heisenberg model.
To second order in the $\epsilon = 4-d$ expansion,
$y_\tau \approx 1.77\pm 0.01$.
The scaling functions $f^\chi_\pm(x)$ obey
\begin{eqnarray}
 f^\chi_+(x) & \to & x^2,\quad  {\rm as }\quad x\to 0; \nonumber \\
 f^\chi_-(x) & \to & {\rm const.},\quad  {\rm as }\quad x\to 0; \\
\label{susc-scaling}
 f^\chi_+(x) & \to & f^\chi_-(x)
\propto x^{1/\delta_H -1},\quad  {\rm as }\quad x\to \infty;
\nonumber
\end{eqnarray}
where $\delta_H^{-1} -1 = 2(1+\eta_H) (y_\tau^H)^{-1} = 1.15\pm 0.01$.
The result (\ref{susc-scaling}) for $f^\chi_+$ implies that the non-linear
susceptibility diverges as the \N/\T\  transition is approached from
from the isotropic (\T) side as
\begin{equation}
{\partial^3M \over \partial H^3} \propto |t|^{\gamma_2},
\hbox{  for  }{H \over H_c|t|^{\phi_\chi}} \ll 1,
\label{non-lin-chi}
\end{equation}
with
\begin{equation}
\gamma_2 = 2(\beta_H - \phi_\chi) = -0.52\pm0.01.
\end{equation}
Recall that in this topologically ordered, but
isotropic phase,
$\bf M$ is always along the applied field $\bf H$.
{From} (\ref{chi}) and (\ref{susc-scaling}), one deduces a non-analytic
(though non-divergent) temperature dependence of the {\em linear}
susceptibility as the \N/\T\  transition is approached from the nematic side,
\begin{equation}
{\partial M \over \partial H} = \chi_I +
{2 \over 3} \Delta \chi = a + b|t|^{2\beta_H},
\label{lin-chi}
\end{equation}
where $a$ and $b$ are constants.
Increasing the magnetic field so that
$H \ll H_c |t|^{\phi_\chi} $, the singular temperature dependence
%${H \over H_c |t|^{\phi_\chi} } \ll 1$,
(\ref{non-lin-chi}) of the non-linear susceptibility becomes a
singular magnetic field dependence
\begin{equation}
{\partial^3M \over \partial H^3} \propto H^{\delta_H^{-1} -3}, \hskip.2cm
{H \over H_c|t|^{\phi_\chi}} \gg 1,
\label{sing-H-dep}
\end{equation}
which is readily seen to cross over smoothly to the low-field
($ {H \over H_c|t|^{\phi_\chi}} \ll 1$) result at $H \sim H_c |t|^{\phi_\chi}$.

\vskip4pt
{\bf 8. Freedericksz instability.}
The Freedericksz instability \cite{de Gennes} provides a means of
measuring the ratio of the anisotropic susceptibility
$\Delta \chi$ to the Frank constants $K_i$.
When a nematic is confined between parallel plates which have been treated
to create boundary conditions favoring a particular orientation of
the director, it will assume that orientation even in the presence of an
applied magnetic field $\bf H$ that favors a different orientation,
as long as that field is sufficiently small.
Usually experimentalists choose this field to be perpendicular
to that favored by boundary conditions.
When the strength of the applied field exceeds a threshold $H_F$, however,
the director rotates away from the alignment favored by the boundaries to
that favored by the field.  This is the Freedericksz instability.
The standard result for the threshold field is
\begin{equation}
H_F = {1 \over L} \sqrt{{K_i \over \Delta \chi}} \times O(1).
\label{FreedH}
\end{equation}
The precise geometry of the experiment determines which Frank $K_i$
appears, as well as the $O(1)$ factor.
Inserting the known scaling forms
\begin{equation}
K_i(t,H) = |t|^{\nu_H} f_\pm^i\left( {H \over
{H_c|t|^{\phi_\chi}}} \right)
\label{Kscale}
\end{equation}
and equation (\ref{chi}) for $\Delta \chi$ into (\ref{FreedH}), we obtain
\begin{equation}
H_F = L^{-1+{\eta_H \over 2}} f_\pm^F(L/\xi_H),
\end{equation}
where the scaling function $f_\pm^F(x)$ has the asymptotic forms
\begin{eqnarray}
 f^F_-(x) & \to & x^{-\eta/2},\quad {\rm as }\quad x\to \infty; \nonumber \\
 f^F_+(x) & \to & c,\quad  {\rm as }\quad x\to 0 \nonumber; \\
 f^F_+(x) & \to & c^\prime e^{- c^{\prime \prime} x},\quad
{\rm as }\quad x\to \infty,
\label{Kasy}
\end{eqnarray}
for some constants $c,c^\prime$ and $c^{\prime \prime}$.
These relations imply that
\begin{equation}
H_F \propto
\left\{ \begin{array}{ll}
|t|^{-{\eta \nu \over 2}}L^{-1} \propto |t|^{-0.007}
& \hbox{  for  } t<0, L \gg \xi; \\
L^{-1+{\eta \over 2}} = L^{(-0.99\pm 0.01)}
& \hbox{  for  } L \ll \xi; \\
e^{-{L \over \xi}} \propto e^{-L|t|^\nu} \propto e^{-L|t|^{0.7}}
& \hbox{  for  } t>0, L \gg \xi.
\end{array}
\right.
\label{HF-scaling}
\end{equation}
Unfortunately, the smallness of $\eta_H$ makes the dependence of $H_F$ on
the temperature (and on the anomalous length scale in the critical regime)
extremely weak in the nematic phase, and hence difficult to discern
experimentally.
In practice, one expects to observe an $H_F$ which is nearly constant with
increasing temperature, until it begins to drop in the unusual stretched
exponential manner indicated in (\ref{HF-scaling}).
The observation of either this stretched exponential or the apparent
temperature independence of $H_F$ below $T_{NT}$ would confirm our theory.

The constancy of $H_F$ below $T_{NT}$ should be contrasted
with the predictions of Landau theory in the case of a weakly first-order
transition.  These predictions are
\begin{eqnarray}
\Delta \chi(T) & \propto & Q_0(T) \propto \sqrt{|t|}, \nonumber \\
K_i(T) & \propto & Q_0^2 \propto |t|.
\end{eqnarray}
Given these scalings, (\ref{FreedH}) would imply that
$H_F \propto |t|^{1/4}$, instead of (\ref{HF-scaling}).

\vskip4pt
{\bf 9 Magnetic susceptibility.}
The elusive \I/\T\  transition can also be detected by magnetic
susceptibility measurements, albeit with more difficulty.  We find that
the non-linear susceptibility has a $|t|^{1-\alpha_I}$ singularity at the
transition:
\begin{equation}
{\partial^3 M \over \partial H^3} = c |t|^{1-\alpha_I} + {\rm analytic}.
\label{chi3-IT}
\end{equation}

The rest of this paper is devoted to deriving these conclusions.
The results quoted above for the specific heat require no further discussion,
as they follow directly and immediately from what has already been shown
(in Paper I) about the partition function, which determines
all thermodynamic functions.
The other results involve various correlation functions.

\section{Polarized Light Scattering Near \I/\T\  Transition}
\label{pol-IT}

Consider first light scattering near the \I/\T\  transition.
In general,\cite{Forster} light scattering in an isotropic material
is given by $I_{xx}({\bf q}) \propto \langle |\epsilon({\bf q})|^2 \rangle$.
This scattering is caused by local fluctuations in the (isotropic)
dielectric constant $\epsilon$ which in turn are caused by
density fluctuations of the various components of the material.
For simplicity we consider here a situation with just one such density $\rho$;
including more does not alter the conclusions.
Expanding the density dependence of $\epsilon$ to linear order about
the average density, we obtain
$I_{xx}({\bf q}) \propto \langle |\delta \rho({\bf q})|^2 \rangle$.
The calculation of the polarized light scattering near the \I/\T\
transition thus reduces to determining the fluctuations of the
(non-ordering) density $\rho$.
To calculate these correlations, we must know how
density fluctuations enter the Hamiltonian.

Recall that the basic Hamiltonian defining our theory (equation
\ref{gauge} of Paper I) is
\begin{equation}
-\beta {\cal H} = J \sum_{\langle i,j \rangle}
U_{ij}\S_i \cdot \S_j +
K \sum_{\{ijkl\}}U_{ij}U_{jk}U_{kl}U_{li},
\label{gauge}
\end{equation}
where the spins $\S_i$ are three-dimensional unit vectors on the sites
of a lattice (cubic for convenience), the variables
$U_{ij} = \pm 1$ are associated with links $(ij)$ between
nearest-neighbor sites, and the second sum in the Hamiltonian runs over
elementary plaquettes $ijkl$.

We showed in Paper I that the spins are irrelevant at small $J$, so
they may be neglected when one incorporates density-fluctuation effects
into an effective renormalized lattice gauge Hamiltonian.
There are no symmetry restrictions
upon the dependence of the Hamiltonian on $\rho$.  For long wavelengths,
the dominant dependence is upon the value of $\rho$ itself,
and spatial gradients can be neglected.
A Hamiltonian with the appropriate symmetry-allowed coupling is
\begin{equation}
{\cal H} = \sum \left\{ V(\rho_\mu(x)) - (K_R + \gamma \delta \rho_\mu(x))
U\right\},
\label{dens-Ham}
\end{equation}
where $U$
abbreviates the renormalized plaquette term from eqn. (\ref{gauge}) above.
(Recall that $U$ measures defect density, so that eq. (\ref{dens-Ham})
describes a coupling between defect density and fluid density.)
In the lattice model, the local density variables $\rho_\mu$ are associated
with the plaquettes, so we denote this dependence simply by the directional
index $\mu$.  (Each plaquette is then labeled by its normal vector.)
$K_R$ is the renormalized defect stiffness obtained after integrating
out the spins as described in Paper I, $\gamma$ is the coupling
between $\rho$ and disclination density, and $\delta \rho_\mu
= \rho_\mu(x) - \rho_0$.

For the moment, all we need to know about
$V(\rho)$ is that it is a local, {\em smooth} function through the
transition, since $\rho$ is non-ordering, and it has a minimum {\em near}
(but not at!) the equilibrium density $\rho_0$.
A duality transformation can be applied to the model implied by Hamiltonian
(\ref{dens-Ham}).  As explained in Paper I,
the duality transformation is completely local in the couplings, so
the partition function becomes
\begin{equation}
Z = \int D\rho \sum_{\{\sigma_\alpha\}}
\exp \biggl\{ \sum_\mu (-V_R(\delta \rho_\mu) + J_\mu(\delta \rho_\mu)
\sigma_\alpha \sigma_{\alpha + \mu}) \biggr\},
\end{equation}
with a ``renormalized'' potential $V_R$ and density-dependent spin coupling
$J$ given by
\begin{eqnarray}
V_R(\rho_\mu) & = & V(\rho_\mu) - (K_R + \gamma \delta \rho_\mu) \nonumber \\
& & \qquad - 1/2 \ln(1-e^{-4K(r_\mu)}) \nonumber \\
J(\delta\rho_\mu) & = & 1/2 \ln \coth(K_R + \gamma \delta \rho_\mu).
\end{eqnarray}

In this model, fluctuations in $\rho$ are driven entirely by those
of the Ising spins that are dual to the gauge field.
More precisely, they are proportional to the energy-energy
correlations of the Ising model obeying the scaling law (\ref{light1}).
To see this, expand $V_R$ to second order and
$J(\delta \rho)$ to first order in powers of $\delta \rho$.
The result for the Hamiltonian is
\begin{eqnarray}
H & \approx & H_{{\rm Ising}}(\rho_0) + h \sum \delta \rho_\mu
-\gamma_R \sum \delta\rho_\mu \sigma_\alpha \sigma_{\alpha + \mu}
\nonumber \\
& + & A/2 \sum (\delta\rho)^2
\end{eqnarray}
with
\begin{equation}
h  = V_R^\prime(\rho_0),\,  A  =  V_R^{\prime \prime}(\rho_0),
\, \hbox{\rm and  }
\gamma_R  =  J^\prime(\rho_0),
\end{equation}
where primes indicate derivatives.
The calculations required to find expressions for
$\langle \delta\rho_\mu \rangle $ and
$\langle \delta\rho_\mu \delta\rho_{\mu^\prime} \rangle $
are easy in this approximation since
they are Gaussian integrals over $\delta\rho$
(fluctuations on different sites are even decoupled).
Specifically,
\begin{equation}
-A \langle \delta\rho_\mu \rangle = h - \gamma_R \langle \sigma_\alpha
\sigma_{\alpha + \mu} \rangle.
\end{equation}
Since this is zero by definition, $h = \gamma_R \langle \sigma_\alpha
\sigma_{\alpha + \mu} \rangle$.  Using this expression for $h$,
\begin{equation}
A^2 \langle \delta\rho_\mu \delta\rho_{\mu^\prime} \rangle =
\gamma_R^2 \langle \sigma_\alpha \sigma_{\alpha + \mu};
\sigma_{\alpha^\prime} \sigma_{\alpha^\prime + \mu^\prime} \rangle
+ A \delta_{\mu \mu^\prime},
\end{equation}
where $\langle x; y \rangle \equiv \langle xy \rangle - \langle x \rangle
\langle y \rangle$ is the truncated correlation function.
Aside from the zero-range second term, the right hand side of this
equation is simply the energy-energy correlation function for the
three-dimensional Ising model,
the well-known behavior of which now implies
the result eq. (\ref{light1}) (Item 2).

\section{Depolarized Scattering Near \N/\T\  Transition}
\label{Depolar-NT}

Next, we turn to depolarized light scattering near the \N/\T\  transition.
The standard form for such scattering in an {\em anisotropic} medium is
\cite{Forster}
\begin{equation}
I_{ij}({\bf q}) \propto \langle \epsilon_{ij}({\bf q})
 \epsilon_{ij}(-{\bf q}) \rangle,
\end{equation}
where $\epsilon_{ij}({\bf q})$ is the Fourier transform of the
position-dependent local dielectric tensor and the summation
convention has been suspended on the right-hand side.

For small nematic order the dielectric tensor can be expanded in the
tensor nematic order parameter
\begin{equation}
Q_{ij}({\bf r}) = S_i({\bf r})S_j({\bf r}) - \delta_{ij}/3
\label{QfromS}
\end{equation}
as
\begin{equation}
\epsilon_{ij} = \epsilon_0 \delta_{ij} + g Q_{ij},
\end{equation}
where $g$ is a microscopic parameter, and is therefore
nonsingular at the transition.

The correlation function
\begin{equation}
C_{ij}({\bf r}) \equiv
\langle \epsilon_{ij}({\bf r}) \epsilon_{ij}({\bf 0}) \rangle
\end{equation}
for $i \not = j = z$ can therefore be expressed in terms of Heisenberg spins
of our model as
\begin{equation}
C_{iz}({\bf r})
\propto \langle \S_i({\bf r}) \S_z({\bf r})
\S_i({\bf 0}) \S_z({\bf 0}) \rangle.
\end{equation}

In Paper I we have shown that near the \N/\T\ transition the system
is described by an effective Heisenberg model, so that
we can apply a renormalization group transformation to relate
$C_{iz}({\bf r})$
to the same function in the Heisenberg model with different parameters:
\begin{eqnarray}
C_{iz}({\bf r},t) & = & b^{-2(d-2+\eta_H)} C_{iz}(b^{-1}{\bf r}, b^{1/\nu_H}t)
\nonumber \\
& = & r^{-2(d-2+\eta_H)} g_{\pm}(r/\xi).
\end{eqnarray}
The second equality follows upon making the choice $b=r/a$,
where $a$ is a microscopic length and
$\xi = \xi_0 |t|^{-\nu_H}$ is the three-dimensional Heisenberg
correlation length, noting the isotropy of the Heisenberg model,
and making the definition
\begin{equation}
g(x) \equiv C_{iz}(a,x^{1/\nu_H}(\xi_0/a)^{1/\nu_H}) a^{2(d-2+\eta_H)}.
\end{equation}
Fourier transformation (and the substitution $d=3$) leads directly to
eq. (\ref{light4}) (Item 4), with
\begin{equation}
f^{NTD}_\pm(q\xi) =  \Gamma \int d^3R {g_\pm(R/q\xi) \over R^{2(1+\eta)}}
e^{iR_z}.
\end{equation}

To complete the description of the critical scattering near the \N/\T\
transition requires matching the scaling form (\ref{light4}) onto known
long-wavelength behavior in the \N\ and \T\ phases to obtain the
asymptotic forms of $f_\pm$.

Consider first the topologically ordered (\T) phase ($t > 0$).
No Goldstone modes are present, since there is no broken symmetry.
The fluctuations of the dielectric tensor (hence light scattering)
must therefore be {\em finite} as $|{\bf q}| \to 0$ at fixed temperature.
Consistency with the scaling law
demands $f_+(x) \propto x^{1-2\eta_H}$ as $x \to 0$.  The $q \to 0$
scattering therefore diverges upon approaching the transition from
the isotropic side:
\begin{equation}
I_{iz} (q \to 0, t ) \propto \xi^{1-2\eta_H} \propto t^{-\nu(1-2\eta_H)},
\: q\xi \ll 1.
\end{equation}
The requirement of finite scattering at fixed $q$ in the critical
regime ($q\xi \gg 1$) implies that $f_+$ and $f_-$ tend to the same constant
as $x \to \infty$;  this means
\begin{equation}
I_{iz}(q) \propto q^{-1+2\eta_H}, \hbox{\rm  for  } q\xi \gg 1.
\end{equation}

On the nematic side of the transition ($t < 0$), scattering is
dominated by Goldstone modes --- fluctuations in the orientation of the
director.  The resulting long-wavelength behaviors of eqs. (\ref{For-1},
\ref{For-2}) are standard results\cite{Forster}.
In these equations, the anisotropy of the dielectric tensor,
$\Delta \epsilon$, is given by
\begin{equation}
\langle \epsilon_{ij} \rangle \equiv \epsilon_0 \delta_{ij}
+ \Delta \epsilon \ \left( n_i n_j -{1 \over 3}\delta_{ij} \right),
\end{equation}
the mean director {\bf n} lies along
${\bf \hat z}$, and ${\bf q} \cdot {\bf \hat y} = 0$.
If the scaling results are to coincide with eqs. (\ref{For-1}, \ref{For-2})
in the $q \to 0$ limit,
then $f_-(x) \propto x^{-(1+2\eta)}$ as $x \to 0$.  As a result,
\begin{eqnarray}
I_{iz}(q) \sim & \xi^{-(1+2\eta_H)} q^{-2} \\
& \propto |t|^{\nu(1+2\eta_H)} q^{-2}, &
\hbox{  as  } q\xi \to 0 \hbox{  and  } t \to 0^-.
\label{again}
\end{eqnarray}
This is the result announced in equation (\ref{light3}) of Item 4.
Implications for the critical behavior of the Frank constants are
also carried by eq. (\ref{again}) combined with
eqs. (\ref{For-1}, \ref{For-2}).
Specifically,
\begin{equation}
{(\Delta \epsilon)^2 \over K_i} \propto |t|^{\nu_H(1+2\eta_H)}, \: i=1,2,3.
\label{eps-and-K}
\end{equation}
Clearly all three Frank constants have the same critical behavior.

We now verify both eq. (\ref{eps-and-K}) and the critical behavior
($K_i \sim |t|^{\nu_H}$) of the Frank constants by an independent derivation.
If we take an expectation value and express $Q_{ij}$ in terms of spins
(see eq. (\ref{QfromS})) we find
\begin{equation}
\langle Q_{ij} \rangle \propto |t|^{2\beta} (n_i n_j -\delta_{ij}/3).
\end{equation}
Comparing with the definition of $\Delta \epsilon$, we deduce $\Delta
\epsilon \propto t^{2\beta_H}$.

Some information about the Frank constants can be obtained from
generalized Josephson relations\cite{Josephson}.
Imagine imposing a twist on the boundary conditions for a nematic
confined to an $L \times L \times L$ volume, and compare to one with
periodic boundary conditions.  The (Frank) free energy difference
is $\Delta F(L) =  (K_2/2) L \theta^2$, where $\theta$ is the twist angle.
{From} scaling, one expects $\Delta F = g(L/\xi)$ near a continuous
transition.  Combining these two equations,
\begin{equation}
K_2 \propto \xi^{-1} \propto |t|^{\nu_H}.
\end{equation}
Very similar arguments that use boundary conditions which impose
a bend or splay distortion imply $K_1$, $K_3 \propto |t|^\nu$ as well.
Gathering these results and the scaling of $\Delta \epsilon$
we find
\begin{equation}
{(\Delta \epsilon)^2 \over K_i} \propto |t|^{4\beta_H - \nu_H}.
\end{equation}
Application of the exact scaling relation
$2 \beta = \nu (d-2+\eta) = \nu (1+\eta)$  (the last equality
following from $d=3$) recovers the result eq. (\ref{eps-and-K}).

\section{Corrections to Scaling for Frank Constants}
\label{corrections}

Let us next consider the corrections to scaling for the
Frank elastic constants.  In the context of an $|{\S}|^4$ soft-spin
theory\cite{Ma}, the starting point is a Landau free energy expressed
in terms of a vector-valued field $\S$ with fluctuating magnitude and
direction.  The term lowest order
in gradients is uniquely of the form $f(|\S|^2) (\nabla \S)^2$, as long
as the invariance under global rotations of $\S$ is unbroken (compare
the discussion in Paper I).  In real nematics, this invariance is broken
and only the invariance under combined identical rotations in internal and
real space survives.

The Landau-Ginsburg free energy then contains three types of gradient terms:
\begin{eqnarray}
{\cal F}_{\mathrm{\scriptstyle grad}} =
K_1 (\nabla \cdot \S)^2 & + &
K_2 |\S \cdot (\nabla \times \S)|^2 +  \nonumber \\
&& K_3 |\S \times (\nabla \times \S)|^2,
\end{eqnarray}
where $K_1$, $K_2$ and $K_3$ are the Frank constants for splay, bend, and
twist, respectively.  In an $\epsilon = 4-d$ expansion, the renormalization
group eigenvalues of these operators at the $n=3$ Wilson-Fisher fixed point
are $\lambda_{2,3} = -2 + {\cal O}(\epsilon)$,
as can be seen by power counting (four powers of $\S$ and two gradients).
Since these eigenvalues are large and negative, the difference between
$K_2$ and $K_3$ vanishes rapidly as the
transition is approached.  The standard form for the correction to
scaling is
\begin{equation}
K_i(t) = K_0 |t|^\nu (1+\Delta_i |t|^{\lambda_i \nu}), \: i=1,2,3,
\end{equation}
with $\lambda_2$ and $\lambda_3$ given above.

The difference between $K_2$ and $K_3$ therefore vanishes rapidly, but $K_1$
is another story.  The eigenvalue of the operator $(\nabla \cdot \S)^2$
is\cite{Aharony}
$\lambda_1 = -\epsilon^2/108 + {\cal O}(\epsilon^3)$, so this Frank
splay constant vanishes very slowly under renormalization.
By way of illustration, for
$\epsilon = 1$ and $\Delta_1 = 1$,  $K_1$ still differs by 50\% from the
value $K_0 |t|^\nu$ taken by $K_2$ and $K_3$ at the ridiculously tiny
reduced temperature
$|t| \approx 2^{-100/\nu} \approx 10^{-43}$!

In fact, $\lambda_1$ is so small that for all reasonable values of
$\Delta_1$, the renormalization group flows of
$\Delta K_1 \equiv  (K_1 - K_{2,3}) $ are presumably dominated by
a non-linear term proportional to $(\Delta K_1)^2$
(which to our knowledge has never been calculated).
The full recursion relation for $\Delta K_1$ is then of the form
\begin{equation}
{d \Delta K_1 \over d \ell } = \lambda_1 \Delta K_1 - c (\Delta K_1)^2,
\label{recur}
\end{equation}
where $c$ is
a constant (again, to our knowledge, never calculated).
Due to the smallness of $\lambda_1$, eq. (\ref{recur}) is valid outside an
extremely narrow range of temperatures in which $\Delta K_1 \ll \lambda_1$.
Assuming $c > 0$ -- {\it i.e.} the fixed point $\Delta K_1 = 0$ is
globally, as well as locally, stable -- and neglecting the linear term,
we conclude that $\Delta K_1$ vanishes as $1/\ell$ for all experimentally
observable temperatures.

If we choose $\ell$ such that $e^\ell \xi = 1$, we find
$\ell \propto |\ln |t| |^{-1}$. This in turn implies that
$(K_1-K_{2,3})/K_{2,3}$ vanishes like $|\ln(|t|)|^{-1}$, as
claimed in Item (5). This dependence is so weak that in many situations
it will simply look as if the correction term is independent of $t$.
In this case, we can write $K_1(t) = K_0 |t|^\nu
(1+\Delta_1)$ for all practical purposes.  That is, $K_1$ will
vanish with the same exponent as $K_2$ and $K_3$, but with a
prefactor of (presumably) order one.
This means that for ${\bf q}\cdot {\bf \hat y} = 0$, $I_{xz}({\bf q})$ will
remain {\em anisotropic}:
\begin{equation}
I_{xz}({\bf q}) = { c (\Delta \epsilon)^2 K_0^{-1} |t|^{- \nu_H} \over
(1+ \Delta_1)q_x^2 + q_z^2 }.
\end{equation}
The singular character of the anisotropy should be unobservable
for all practical purposes.  $I_{yz}({\bf q})$ is still expected to become
isotropic.  Note that the anisotropy of $I_{xz}$ does not alter the
scaling law $I_{xz} \propto |t|^{\nu(1+2\eta)}q^{-2}$, but rather
introduces a (nonsingular) anisotropic prefactor.

\section{Magnetic Susceptibility Near \N/\T\  Transition}
\label{magnetic-susc}

The application of a magnetic field $\bf H$ adds to the Hamiltonian a term
\begin{equation}
{\cal H}_{\mag} = - \Delta \chi_0 \sum_i ({\bf H}\cdot \S_i)^2,
\label{mag-e}
\end{equation}
where the {\em molecular} diamagnetic anisotropy $\Delta \chi_0$ is
assumed to be a smooth function of temperature, in the spirit of modern
critical phenomenology.
This standard expression eq. (\ref{mag-e}) for the nematic magnetic energy
is the lowest order\cite{loworder} term in ${\bf H}$ which respects
the gauge symmetry ($\S_i \to \sigma_i\S_i$, $U_{ij} \to \sigma_i
\sigma_j U_{ij}$).

The magnetization $\bf M$ induced by the applied magnetic field can
be deduced {\it via} the standard thermodynamic relation
\begin{equation}
M_\alpha  =  {1 \over N} {\partial F \over \partial H_\alpha}
=  2 \Delta \chi_0 \langle S_i^\alpha S_i^\beta \rangle H_\beta.
\end{equation}

Now we know that
\begin{equation}
\langle S_i^\alpha S_i^\beta \rangle = Q_0 (n_\alpha n_\beta -
\delta_{\alpha \beta}/3) + \delta_{\alpha \beta}/3,
\label{StoQ}
\end{equation}
where $Q_0(T,H)$ is the magnitude of the nematic tensor order parameter.
This immediately implies eq. (\ref{magnetic-resp}), with
\begin{equation}
\chi_I = (2/3) \Delta \chi_0,
\label{chiI}
\end{equation}
and
\begin{equation}
\Delta \chi = 2 \Delta \chi_0 Q_0(T,H).
\label{delchi}
\end{equation}
Equation (\ref{chiI}) assures that $\chi_I$ is perfectly smooth
through $T_{NT}$, since $\Delta \chi_0$ is; this follows from the
fact that $|\S|^2 = 1$, regardless of temperature.  Hence, taking the trace
over $\alpha$ and $\beta$ in equation (\ref{StoQ}) fixes the coefficient
of $\delta_{\alpha \beta}$ at {\em exactly} 1/3, independent of temperature
(the first term is traceless by construction and does not affect
this result).  This coefficient, in turn, leads to the {\em exact}
relation (\ref{chiI}) between $\chi_I$ and $\Delta \chi_0$, which in turn
implies analyticity of $\chi_I$.

We must still determine the scaling of $Q_0(T,H)$.
Since $Q_0$ involves correlations of two $\S$ fields, its renormalization
group eigenvalue is simply (see the appendix)
$-2\beta_H/\nu_H$.  Therefore, the renormalization
group transformation near the $d=3$ Heisenberg critical fixed point can be
used to relate $Q_0(T,H)$ to $Q_0$ at rescaled temperature and field as
\begin{equation}
Q_0(T,H) = b^{- 2\beta_H / \nu_H} Q_0(b^{1/ \nu_H}t,b^{y_h}H),
\label{Qscale}
\end{equation}
where $y_h$ is the renormalization group eigenvalue of $H$.
This eigenvalue can be determined by rewriting the magnetic Hamiltonian as
\begin{equation}
{\cal H}_{\mag} =
\Delta \chi_0 \sum_i H_\alpha H_\beta (S_i^\alpha S_i^\beta
-1/3 |\S|^2 \delta_{\alpha \beta})
\label{mag-e-2}
\end{equation}
which differs from (\ref{mag-e}) only by a constant (independent of $\S$),
owing to $|\S|^2 =1$.  In this form, ${\cal H}_{\mag}$ is simply
the ``spin tensor interaction'' of reference \onlinecite{Ma}.
The renormalization group eigenvalue of this perturbation has the $\epsilon
=4-d$ expansion\cite{Aharony}
\begin{eqnarray}
\label{yt-eig}
y_\tau & = & 2\left(1-{\epsilon \over n+8} +
{n^2-18n-88 \over 4(n+8)^2}\epsilon^2\right)
+ {\cal O}(\epsilon^3) \nonumber \\
& \simeq & 1.77\pm0.01,
\end{eqnarray}
where in the second, approximate, equality we have set $d=3$ and $n=3$.  The
quoted error is typical of second order in $\epsilon$ expressions for
$d=3$ $O(n)$ models.  Since the coefficient of the spin tensor perturbation
(\ref{mag-e-2}) is $H^2$, the renormalization group eigenvalue
of $H$ is $y_h = y_\tau/2 \simeq
0.89\pm0.01$.  With this eigenvalue in hand, we can now obtain the scaling of
$Q_0$.
Choosing $b=|t|^{-\nu_H}$ in eq. (\ref{Qscale}), and
using (\ref{delchi}) to relate $Q_0$ to $\Delta \chi$,
we immediately obtain eq.(\ref{chi}) for $\Delta \chi$ (Item 7), with
\begin{equation}
f_I(x) = Q_0(t=\pm 1,H=x).
\label{scale-fun}
\end{equation}
To obtain the asymptotic form (\ref{susc-scaling}) of $f_+$, note that
deep within the disordered phase
the susceptibility of $Q_0$ to $H^2$ (the coefficient of the
magnetic perturbation) should  be finite. Hence $Q_0 \propto H^2$ in
that regime. ($Q_0$ of course vanishes as $H \to 0$ in
the disordered phase, so this $H^2$ term is the leading order term.)
In the ordered phase, $Q_0$ remains non-zero as $H \to 0$, which
implies the asymptotic form (\ref{susc-scaling}) for $f_-$.  Finally,
requiring that
$Q_0(t,H)$ remain finite and non-zero as $t \to 0$ for fixed $H$ implies
the final asymptotic form in eq. (\ref{susc-scaling}) for $f_\pm(x)$
in the critical ($x \to \infty$) regime.

All the other results eqs. (\ref{non-lin-chi}-\ref{sing-H-dep})
quoted for the magnetic susceptibility now follow
straightforwardly from the scaling law eq. (\ref{chi}) and the asymptotic
expression in eq. (\ref{susc-scaling}) for $f_+$.

\section{Freedericksz Instability}
\label{Freedericksz}

Regarding the Freedericksz instability, we need only derive the scaling
form (\ref{Kscale}) for $K_i$, and the asymptotic forms (\ref{Kasy}).
The former can be done in a manner precisely parallel to the argument
given for $Q_0$.  The difference is that the Frank constants have
renormalization group eigenvalue 1, as can be seen by requiring that the
zero-field Josephson relation $K_i \propto |t|^\nu$ holds.

The first asymptotic form in eq. (\ref{Kasy}) then follows from the standard
result $H_F \propto L^{-1}$, which holds when nematic order is well-developed.
The second expression in eq. (\ref{Kasy}) follows by requiring that $H_F$
remain
finite and non-zero in the critical regime.
In the disordered phase, boundary effects surely fall off
exponentially with the distance from the boundary for distances large
compared to the correlation length.  This gives the last
(exponential) form in eq. (\ref{Kasy}).

\section{Magnetic Susceptibility Near \I/\T\  Transition}
\label{susceptibility-IT}

Finally, we consider the magnetic susceptibility and {\em depolarized}
light scattering near the \I/\T\  transition.
All of the manipulations up to equation (\ref{delchi}) are the same here as
near the \N/\T\  transition.  The differences all arise in the scaling
behavior of $Q_0(t,H)$.  This can be obtained by noting that, since
there is no {\em spontaneous} nematic ordering anywhere near the \I/\T\
transition (except at the multicritical point),
the nematic director will necessarily be along the applied
field $\bf H$. (Without loss of generality, we will consider $\bf H$
along the z-direction.)  In that case,
\begin{equation}
\langle S_i^\alpha S_i^\beta \rangle = Q_0 (\delta_{\alpha z} \delta_{\beta z}
-\delta_{\alpha \beta}/3) + \delta_{\alpha \beta}/3,
\end{equation}
whence
\begin{equation}
Q_0 = {3 \over 2}[ \langle (S_i^z)^2 \rangle - 1/3].
\label{Q0}
\end{equation}

Since the magnetic Hamiltonian can be rewritten
\begin{equation}
{\cal H}_{\mag} = \Delta \chi_0 H^2 \sum_i [(S_i^z)^2 - 1/3]
\end{equation}
by adding a trivial constant, it follows that
\begin{equation}
Q_0 = -{3 \over 2N} {\partial F \over \partial (h^2)}
\label{Q0cis}
\end{equation}
where $N$ is the number of sites in the system,
and $h = H \sqrt{{\Delta \chi_0/k_{_B} T}}$.

Using (\ref{Q0cis}), we can obtain $Q_0$ by familiar manipulations
of the partition function.
When ${\cal H}_{\mag}$ is added to the original gauge theory Hamiltonian,
the resulting partition function is
\begin{eqnarray}
Z & = & \int \prod_i d\Omega_i \sum_{\{U\}}
\exp \biggl\{ K\sum U(P) +  \nonumber \\
&&  J \sum U_{ij} \S_i \cdot \S_j
 + h^2\sum_i[(S_i^z)^2-1/3] \biggr\}.
\end{eqnarray}
As before, we perform a ``polymer'' expansion of the terms
$e^{(JU_{ij}\S_i \cdot \S_j)}$ in the partition function.  Extracting
a factor
\begin{equation}
Z_D = \langle e^{ h^2\sum_i[(S_i^z)^2-1/3]} \rangle_{_S},
\end{equation}
which represents completely decoupled spins ($\langle \cdot \rangle_{_S}$
indicates normalized expectation with completely uniform spin distribution
on the unit sphere),
we rewrite the partition function as
\begin{eqnarray}
Z & = & Z_D Z_G {\sum_{\{C_n\}} }^\prime
\left\langle  \prod_m W(C_m) \right\rangle_{_G}
\nonumber \\
& \prod_m &
{
\langle \prod_{(ij)\in C_m} (J \S_i \cdot \S_j) \prod e^{h^2[(S_i^z)^2-1/3]}
\rangle_{_S}
\over
\langle \prod_{(ij)\in C_m}
\prod e^{h^2[(S_i^z)^2-1/3]} \rangle_{_S}
 }.
\label{poly-mag}
\end{eqnarray}
This is a sum over disconnected polymers $C_n$, with factors of
$e^{h^2[(S_i^z)^2-1/3]}$ decorating the polymers; the same factors for
sites not on polymers have been absorbed into the $Z_D$ out front, which
requires the denominator to avoid double counting.
$Z_D$ is clearly a smooth function of $H$.
Near the \I/\T\  transition the singular contributions to $Q_0$ come from
the fluctuations of the $U_{ij}$'s.
For small $J$, it is safe to expand in $J$.
The calculation of the effective gauge coupling is similar to that in
Paper I (section VII D), and involves resumming the leading order
contributions, {\it i.e.,} the polymers consisting of single disconnected
plaquettes.  The result is
\begin{equation}
K_{\eff} = K + J^4 {
\langle \prod (\S_i \cdot \S_j) \prod
e^{h^2[(S_i^z)^2-{1 \over 3}]} \rangle_S
\over
\langle \prod e^{h^2[(S_i^z)^2-{1 \over 3}]} \rangle_S
}
 + {\cal O}(J^6),
\label{Keff-mag}
\end{equation}

To calculate the lowest order contributions in $H$ and $J$ from
eq. (\ref{Keff-mag}), note that the denominator can be replaced by unity.
In the numerator, the product of $\S_i \cdot \S_j$ terms only
depends upon the {\em relative} orientation of the spins in the polymer.
For the order $H^2$ term,
the absolute orientation of only one spin matters.
Integrating over the orientation of this spin first, with relative
orientations fixed, and then over the relative orientations, the
result is zero since $\langle (S_i^z)^2 \rangle_{_S} = 1/3$.
Thus,\cite{keff} the expansion starts at order $H^4$.

The coefficient of $H^4$ is easily calculated to
lowest order in $J$.  Eq. (\ref{Keff-mag}) becomes
\begin{equation}
K_{\eff} = K + {176 \over 6075} J^4 H^4 + \cdots.
\label{Keff-numbers}
\end{equation}
The most important consequence of this analysis is that the generic
singularity of the magnetic susceptibility near the \I/\T\  transition
has the $1-\alpha_I$ singularity reported in eq. (\ref{chi3-IT}) (Item 9).

All of the critical behavior of $Q_0$ can be
obtained from the field dependence of $K_{\eff}$ {\it via}
equation (\ref{Q0cis}), which gives
\begin{eqnarray}
Q_0 & = &
      -{3 \over 2N} {\partial F \over \partial K_{\eff}}
           {\partial K_{\eff} \over \partial (h^2)} \nonumber \\
    & = & -{3 h^2\over N}\left(k_{_B} T \over \Delta \chi_0\right)^2 C(J)
{\partial F \over \partial K_{\eff}} \nonumber \\
     & = &  (a+b|t|^{1-\alpha}) H^2.
\label{Q0I/T}
\end{eqnarray}
where in the last equality, we have inserted the known singularity of
$F$ at the (zero-field) \I/\T\  transition, as well as the fact that
$\partial F/\partial K_{\eff} \propto \partial F/\partial T$.
As a bonus, (\ref{Q0I/T}) also tells us the field dependence of the
transition temperature $T_{IT}$:
\begin{equation}
T_{IT}(H) = T_{IT}(0) + {\rm const.}\times H^4.
\end{equation}

\section{Depolarized Scattering Near \I/\T\  Transition}
\label{depolar-IT}

This hard-won knowledge of the singular behavior of $Q_0$  near
$T_{IT}$ enables us also to determine the critical behavior of the
depolarized light scattering $I_{xz}({\bf q},t)$ near $T_{IT}$ (Item 3).
As shown earlier, its Fourier transform $I_{xz}({\bf r},t)$ obeys
\begin{equation}
I_{ij}({\bf q}) \propto \langle Q_{ij}({\bf r}) Q_{ij}({\bf 0})
\rangle \equiv C_{ij}({\bf r}).
\label{ft-of-C}
\end{equation}

We can now obtain the behavior of $C_{ij}({\bf r})$ from scaling:
\begin{equation}
C_{ij}({\bf r},t) = b^{2X_Q} C_{ij}(b^{-1}r,b^{1 / \nu} t),
\label{Cij-scaling}
\end{equation}
where $X_Q$ is the renormalization group eigenvalue of $Q_{ij}$
and $b$ the rescaling factor.
We can extract $X_Q$ from the results just obtained for the
singular part of the non-linear susceptibility.  Scaling implies
\begin{equation}
Q_0^{\rm sing}(t,h) = b^{X_Q} Q_0^{\rm sing}(b^{1/\nu}t,b^{\lambda_h}h),
\end{equation}
where $Q_0^{\rm sing}$ is the singular part of the magnitude of the
nematic order, and $\lambda_h$ is the (as yet unknown) renormalization group
eigenvalue of $H$ at the Ising gauge critical point.  Choosing $b=|t|^{-\nu}$,
we obtain
\begin{equation}
Q_0^{\rm sing}(t,h) = |t|^{-\nu X_Q}
f^{ITD}\left({h \over |t|^{\nu \lambda_h}}\right),
\end{equation}
where
\begin{equation}
f^{ITD}(x) \equiv Q_0^{\rm sing}(1,x).
\end{equation}
For this to be consistent with our earlier result (\ref{Q0I/T}) for $Q_0$,
we must have $f(x) \propto x^2$ at small $x$ and also
$|t|^{-\nu(X_Q+2\lambda_h)} = |t|^{1-\alpha}$,
which implies
\begin{equation}
X_Q + 2 \lambda_h = {\alpha-1 \over \nu} .
\end{equation}
To obtain $X_Q$, we still need $\lambda_h$.  But since $h^2$ couples directly
to $Q_0$ ({\it i.e.}, to $S_z^2-1/3$, whose average is $Q_0$),
we know from scaling that $2 \lambda_h = d+X_Q$.
Making this substitution, we find (with $d$ set to 3)
\begin{equation}
X_Q = {1 \over 2} \left( {\alpha -1 \over \nu} -3 \right).
\end{equation}
Using this eigenvalue in (\ref{Cij-scaling}),
and choosing $b = |t|^{-\nu}$, we obtain
the scaling form
\begin{equation}
C_{ij}(r,t) = |t|^{3\nu+1-\alpha} f_{ij}(r/\xi),
\label{Cij-again}
\end{equation}
where $\xi \propto |t|^{-\nu}$ is the Ising transition correlation length.
As the critical point is approached ($t \to 0$),
$C_{ij}(r)$ remains finite but non-zero at fixed $r$, so that the
function $f_{ij}$ must scale in such a way as to offset the $t$ dependence
of the prefactor in (\ref{Cij-again}). This gives the result
\begin{equation}
C_{ij}(r,t=0) \propto r^{(\alpha -1)/ \nu -3},
\end{equation}
which, when inserted back into (\ref{ft-of-C}), implies
\begin{equation}
I_{xz}(q,t=0)  \propto  \int {
e^{i{\bf q}\cdot {\bf r}}
d^3r \over r^{3-(\alpha-1)/ \nu} }
 =  a_2 + b_2 |{\bf q}|^{(1-\alpha)/ \nu},
\end{equation}
where $a_2$ and $b_2$ are constants and the exponent
$(1-\alpha)/\nu$ has the numerical value 1.44.
Note that the scattering
remains finite at $|{\bf q}| = 0$, and lacks even a cusp since the
exponent $q$ is greater than one;
two derivatives are needed to get a divergence
($\partial^2 I_{xz} / \partial q^2 \sim |{\bf q}|^{-0.56}$).

Finally, we argue that
a lack of microscopic head-tail symmetry will not alter these results
near the \I/\T\ transition.  It suffices to show that the
leading order contribution to $K_{\eff}$ remains ${\cal O}(H^4)$,
even in that case.
The $H^4$ term is non-vanishing (even a system lacking head-tail
symmetry will still have a $\Delta \chi_0 \sum ({\bf H}\cdot{\bf S}_i)^2$
term in its magnetic energy), and
we need show only that lower order terms continue to vanish.
This is an immediate consequence of the absence of
nematic order, even in the presence of such a microscopic
symmetry breaking.
No analytic scalar (rotationally invariant) function can be made which is
linear or cubic in ${\bf H}$, and the ${\bf H}^2$ term vanishes by
equation (\ref{Q0cis}).

\section{Conclusion}

In this paper, we have analyzed critical behavior of a variety of
quantities at the transitions into the new, topologically ordered
phase of nematics predicted in our gauge model\cite{part1}.
Characteristic signatures were found for several standard experimental
probes of soft condensed-matter, notably specific heat, and optical
and magnetic response.
We hope that these behaviors will be sought in the laboratory
in the near future.

\begin{acknowledgements}
JT thanks D. Roux for many discussions of his experiment, J. Prost for
pointing out the possible connection of those experiments to this work, and
for hospitality while various portions of this work were underway,
the Aspen Center for Physics, the CNRS Paul Pascal (Bordeaux, France), the
Isaac Newton Institute for Mathematical Sciences (University of Cambridge, UK)
and the Institute for Theoretical Physics of U.C. Santa Barbara
(and their NSF Grant PHY89-04035).
DSR thanks V. Luby for interesting conversations and acknowledges grant
support from NSF PYI91-57414 and the Sloan Foundation.
\end{acknowledgements}

\section{Appendix}
In this appendix we calculate the scaling dimension of $Q_0(T)$.
Writing the spin as $\S = M(t) \hat{\bf z} + \delta \S$, $M$ being the
magnetization arbitrarily assigned to point along $\hat{\bf z}$,
\begin{equation}
\langle Q_{ij} \rangle
= M^2(t) \delta_{iz} \delta_{jz} + \langle \delta S_i
\delta S_j - \delta_{ij}/3 \rangle.
\label{expect1}
\end{equation}
The second term is the expectation valud of the connected piece of the
operator $\tau = S_i S_j - (1/3)\delta_{ij}$.  This expectation
vanishes as $t^{\beta_\tau}$ as $t \to 0$.  By general renormalization
group theory\cite{Ma},
\begin{equation}
\beta_\tau = \nu (d-y_\tau),
\end{equation}
where $y_\tau$ is the RG eigenvalue of the operator $\tau$ at the
relevant (i.e. $n=d=3$ Wilson-Fisher) fixed point.

Thus, the second term in (\ref{expect1}) is asymptotically dominated by the
first as $t \to 0$ provided
$2 \beta = \nu (d+\eta-2) < \beta_\tau$,
or equivalently, if and only if
\begin{equation}
2 \beta - \beta_\tau = \nu (y_\tau +\eta-2) < 0.
\end{equation}
The epsilon expansion for $y_\tau$ can be found in Ma's book
(ref. \onlinecite{Ma}, pg. 355), and is
\begin{equation}
y_\tau = 2 \left( 1 - {\epsilon \over n+8}
+ {n^2-18n-88 \over 4 (n+8)^3} \epsilon^2 \right) + {\cal O}(\epsilon^3),
\end{equation}
which, combined with $\eta = {n+2 \over 2(n+8)} \epsilon^2 +
{\cal O}(\epsilon^3)$ at $n=3$ gives
\begin{equation}
2 \beta - \beta_\tau = \nu \left( -{2\epsilon \over 11} -
{23 \epsilon^2 \over 5724} \right) +
{\cal O}(\epsilon^3),
\end{equation}
which is indeed negative, as desired.  The expectation (\ref{expect1})
is dominated by the connected piece and scales with an exponent $2 \beta$.

\end{document}